%% file: main.tex
\documentclass[conference]{IEEEtran}
\usepackage{multirow}
\usepackage{amsmath,amsfonts}
\usepackage{algorithmic}
\usepackage{algorithm}
\usepackage{array}
\usepackage[caption=false,font=normalsize,labelfont=sf,textfont=sf]{subfig}
\usepackage{textcomp}
\usepackage{stfloats}
\usepackage{booktabs}
\usepackage{url}
\usepackage{verbatim}
\usepackage{graphicx}
\usepackage{cite}
\usepackage{amsmath}
\usepackage{amsfonts}
\usepackage{graphicx}
\usepackage[table,xcdraw]{xcolor}
\usepackage[printonlyused, nolist]{acronym}
\usepackage{float} 
\usepackage{tikz}

\newcommand\copyrighttext{%
  \footnotesize \textcopyright \the\year{} IEEE. Personal use of this material is permitted. Permission from IEEE must be obtained for all other uses, including reprinting/republishing this material for advertising or promotional purposes, collecting new collected works for resale or redistribution to servers or lists, or reuse of any copyrighted component of this work in other works.}

\newcommand\copyrightnotice{%
\begin{tikzpicture}[remember picture,overlay]
\node[anchor=north,yshift=0pt] at (current page.north) {\fbox{\parbox{\dimexpr0.9\textwidth-\fboxsep-\fboxrule\relax}{\copyrighttext}}};
\end{tikzpicture}%
}

\begin{document}

\title{HELENA: High-Efficiency Learning-based channel Estimation using dual Neural Attention}

\author{Miguel Camelo Botero, Esra Aycan Beyaz{\i}t, Nina Slamnik-Kriještorac, Johann M. Marquez-Barja \\
    University of Antwerp - imec, IDLab, Antwerp, Belgium
}

\maketitle

\copyrightnotice 

\begin{abstract}
Accurate channel estimation is critical for high-performance Orthogonal Frequency-Division Multiplexing (OFDM) systems, particularly at low signal-to-noise ratios and under stringent latency constraints. This paper presents \ac{HELENA}, a compact deep learning model that combines a lightweight convolutional backbone with two efficient attention mechanisms: patch-wise multi-head self-attention for capturing global dependencies and a squeeze-and-excitation block for local feature refinement. Compared to CEViT, a state-of-the-art vision transformer-based estimator, HELENA reduces inference time by 45.0\% (0.175ms vs. 0.318ms), achieves comparable accuracy (-16.78dB vs. -17.30dB), and requires $8\times$ fewer parameters (0.11M vs.0.88M), demonstrating its suitability for low-latency, real-time OFDM-based wireless systems.
\end{abstract}


\input{acronyms}
\input{sections/section1}
\input{sections/section2}
\input{sections/section3}
\input{sections/section4}
\input{sections/section5}
\section*{Acknowledgments}
This research was supported by the 6G-TWIN project under the SNS JU Horizon Europe program with Grant Agreement No. 101136314. This research is partially funded by the imec.icon project Rapidness, which is co-financed by imec and Flanders Innovation and Entrepreneurship under project nr HBC.2024.0772. The 6G-TWIN support relates to energy-efficient data-driven functional models for Network Digital Twins, while RAPIDNESS supported the development of a hardware-efficient, low-latency learning-based channel estimation model.
\bibliographystyle{IEEEtran}
\bibliography{main.bib}

\end{document}

%% file: acronyms.tex
\begin{acronym}
    \acro{AI}{Artificial Intelligence}
    \acro{ML}{Machine Learning}
    \acro{CNN}{Convolutional Neural Network}
    \acro{FC}{Fully Connected Layers}
    \acro{5G-NR}{5G New Radio}
    \acro{5G}{fifth-generation}
    \acro{DL}{Deep Learning}
    \acro{ReLU}{Rectified Linear Unit}
    \acro{DNN}{Deep Neural Network}
    \acro{ONNX}{Open Neural Network Exchange}
    \acro{GPU}{Graphics Processing Unit}
    \acro{CPU}{Central Processing Unit}
    \acro{SRCNN}{Super Resolution Convolutional Neural Network}
    \acro{EDSR}{Enhanced Deep Super-Resolution}
    \acro{ViT}{Vision Transformer}
    \acro{SR}{Super-resolution}
    \acro{IR}{Image Restoration}
    \acro{LR}{Low-Resolution}
    \acro{LS}{Least Squares}
    \acro{LI}{Linear Interpolation}
    \acro{HR}{High-Resolution}
    \acro{MDSR}{Multi-Scale Approach}
    \acro{MAE}{Mean Absolute Error}
    \acro{CE}{Channel Estimation}
    \acro{NMSE}{Normalized Mean Squared Error}
    \acro{MSE}{Mean Squared Error}
    \acro{MHSA}{Multi-Head Self-Attention}
    \acro{BER}{Bit Error Rate}
    \acro{CSI}{Channel State Information}
    \acro{3GPP}{3rd Generation Partnership Project}
    \acro{OFDM}{Orthogonal Frequency-Division Multiplexing}
    \acro{SNR}{Signal-to-Noise Ratio}
    \acro{AWGN}{Additive White Gaussian Noise}
    \acro{SISO}{Single Input Single Output}
    \acro{I/Q}{In-Phase and Quadrature Components}
    \acro{1D}{One-Dimensional}
    \acro{2D}{Two-Dimensional}
    \acro{AR}{Augmented Reality}
    \acro{VR}{Virtual Reality}
    \acro{UHD}{Ultra-High-Definition}
    \acro{TDL}{Tapped Delay Line}
    \acro{TTI}{Transmission Time Interval}
    \acro{ReLU}{Rectified Linear Unit}
    \acro{DnCNN}{Denoising Convolutional Neural Network}
    \acro{Conv}{Convolutional}
    \acro{NR}{New Radio}
    \acro{MMSE}{Minimum Mean Square Error}
    \acro{DRN}{Deep Residual Networks}
    \acro{ResNet}{Residual Network}
    \acro{DNN}{Deep Neural Network}
    \acro{RB}{Resource Block}
    \acro{SCS}{Subcarrier Spacing}
    \acro{CP}{Cyclic Prefix}
    \acro{PDSCH}{Physical Downlink Shared Channel}
    \acro{SR-Net}{Super-Resolution Network}
    \acro{CIR}{Channel Impulse Response} 
    \acro{SE}{Squeeze-and-Excitation}
    \acro{ECA}{Efficient Channel Attention}
    \acro{GAP}{Global Average Pooling}
    \acro{CSI}{Channel State Information}
    \acro{SRDnNet}{Super Resolution De-noising Convolutional Neural Network}
    \acro{ReEsNet}{Residual channel Estimation Network}
    \acro{UE}{User Equipment}
    \acro{IQ}{In-phase an Quadrature}
    \acro{HELENA}{High-Efficiency Learning-based channel Estimation using dual Neural Attention}
    \acro{FLOPS}{Floating-point operations per second}
    \acro{Conv2D}{2D Convolutional}
    \acro{FC}{Fully Connected}
    \acro{MBConv}{Mobile Inverted Bottleneck Convolution}
    \acro{HARQ}{Hybrid Automatic Repeat Request}
    \acro{FPGA}{Field-Programmable Gate Array}
    \acro{LSiDNN}{LS-augmented interpolated Deep Neural
Network}
    \acro{AttRNet}{Attention mechanism and Residual
Network}
    \acro{CEViT}{Channel Estimator Vision Transformer}
    \acro{OTFS}{Orthogonal Time Frequency Space}
    \acro{UAV}{Unmanned Aerial Vehicle}
    \acro{LoS}{Line-of-Sight}
    \acro{NLoS}{Non-Line-of-Sight}
    \acro{AI}{Artificial Intelligence}
    \acro{NDT}{Network Digital Twin}
    \acro{MIMO}{Multiple-Input Multiple-Output}
    \acro{TDL}{Tapped Delay Line}
    \acro{dB}{Decibels}
    \acro{ProEsNet}{Progressive Estimation Network}
    \acro{EPformer}{Efficient Parallel Transformer}
\end{acronym}

%% file: sections/section1.tex
\bstctlcite{IEEEexample:BSTcontrol}
\acresetall
\section{Introduction}
\IEEEPARstart{A}ccurate estimation of \ac{CSI} is crucial for the effectiveness of \ac{OFDM}-based wireless communication systems, such as \ac{5G-NR}, as it enables optimal resource allocation, beamforming, and adaptive modulation, key factors that directly influence system capacity and reliability. \ac{CE}, which refers to estimating \ac{CSI} from received and reference signals (e.g., pilots), faces practical limitations: \ac{LS} performs poorly in noisy or high-mobility scenarios, while \ac{MMSE} depends on unavailable channel statistics and incurs high computational complexity, restricting its deployment in real-time wireless systems.


To address these challenges, recent research has investigated both model-based and data-driven approaches for improving \ac{CE} under realistic 5G and beyond conditions. Among model-based techniques, several recent methods have demonstrated competitive performance by leveraging signal structure and auxiliary information, such as pilot-free estimation for high-Doppler \ac{OTFS} scenarios~\cite{QingTVT2025} and sensing-assisted denoising of \ac{LS} estimates~\cite{QingJIOT2024}. In parallel, \ac{DL}-based methods have emerged as powerful tools for enhancing \ac{CE} by learning complex propagation patterns directly from data. These approaches lay the foundation for \ac{AI}-native physical layer design~\cite{Hoydis2021} and the development of the next generation of intelligent radios~\cite{Camelo2020}, which is the focus of this paper.

Architectures such as ChannelNet~\cite{soltani2019deep}, \ac{EDSR}\cite{maruyama2021}, \ac{AttRNet}\cite{Gao2025}, \ac{ProEsNet}\cite{Zhang2024}, and the more recent Transformer-based \ac{EPformer}\cite{Guo2024} and \ac{CEViT}~\cite{Liu2024} achieve superior estimation accuracy over conventional methods. However, their high computational complexity limits their deployment in real-time 5G systems, which are constrained by strict latency, power, and hardware budgets~\cite{Damjancevic2021}. This has led to the development of lightweight architectures such as \ac{LSiDNN}~\cite{Sharma2024}, which aim to balance accuracy with computational efficiency. However, these models tend to achieve lower complexity at the expense of estimation accuracy or real-time performance.

To address these limitations, we introduce \textbf{\ac{HELENA}}, a lightweight hybrid \ac{DL} architecture for pilot-based \ac{CE} in \ac{OFDM} systems without interpolation. It combines convolutional feature extraction with dual attention: \ac{Conv2D} layers for local features, patch-wise \ac{MHSA} for global context, and a post-attention \ac{SE} block for channel refinement. A linear reconstruction head and residual connection preserve structural information and support convergence, especially under high-\ac{SNR} conditions. In this paper, we show that \ac{HELENA} (i) achieves better accuracy–efficiency trade-offs than several state-of-the-art \ac{DL} models, (ii) reveals that computational cost does not always correlate with inference time, and (iii) is fully reproducible via open-source code and data\footnote{\url{https://github.com/miguelhdo/HELENA_Channel_Estimation}}.


%% file: sections/section2.tex
\section{System Model and Problem Statement}

We consider a downlink \ac{SISO} \ac{OFDM} system compliant with 5G-NR. The main notation used throughout the paper is summarized in Table~\ref{tab:notation} for clarity.

\begin{table}[t]
\caption{Summary of Notation}
\label{tab:notation}
\centering
\begin{tabular}{ll}
\toprule
\textbf{Symbol} & \textbf{Description} \\
\midrule
$\mathbf{H}$ & True channel matrix over subcarriers and OFDM symbols \\
$\hat{\mathbf{H}}$ & Estimated (reconstructed) channel matrix \\
$\mathbf{H}_{\text{LR}}$ & Low-resolution channel input (pilot-based or interpolated) \\
$\mathbf{H}_p$ & Pilot-based channel observations \\
$\mathcal{P}$ & Set of pilot positions in the time-frequency grid \\
$f_{\Theta}$ & Deep learning-based channel estimation model \\
$\Theta$ & Learnable model parameters \\
$\mathbf{X}$ & Transmitted OFDM signal matrix \\
$\mathbf{Y}$ & Received OFDM signal matrix \\
$\mathbf{Z}$ & Additive white Gaussian noise matrix \\
$X_{i,k}$ & Transmitted symbol at subcarrier $i$ and OFDM symbol $k$ \\
$Y_{i,k}$ & Received symbol at subcarrier $i$ and OFDM symbol $k$ \\
$H_{i,k}$ & Channel coefficient at subcarrier $i$ and OFDM symbol $k$ \\
$N_S$ & Number of subcarriers \\
$N_D$ & Number of OFDM symbols \\
$p$ & Patch height along the frequency dimension \\
$N$ & Number of patches (tokens), $N = N_S / p$ \\
$\sigma^2$ & Noise variance \\
$\mathrm{LS}$ & Least Squares channel estimation \\
$\mathrm{MMSE}$ & Minimum Mean-Square Error estimation \\
$\mathrm{LI}$ & Linear interpolation \\
$\mathrm{SR}$ & Super-resolution-based channel reconstruction \\
$\|\cdot\|_F$ & Frobenius norm \\
$T_{\mathrm{inf}}(f_\Theta)$ & Inference time of model $f_\Theta$ \\
$T_{\max}$ & Maximum allowable inference latency \\
$\mathrm{MSE}$ & Mean Squared Error \\
$\mathrm{NMSE}$ & Normalized Mean Squared Error \\
$\mathbb{E}[\cdot]$ & Expectation operator \\
\bottomrule
\end{tabular}
\end{table}

The received signal at subcarrier \( i \) and \ac{OFDM} symbol \( k \) is given by
\begin{equation}
Y_{i,k} = H_{i,k}X_{i,k} + Z_{i,k},
\end{equation}
where \( X_{i,k} \) is the transmitted symbol, \( H_{i,k} \) the channel coefficient, and \( Z_{i,k} \sim \mathcal{CN}(0, \sigma^2) \) represents additive white Gaussian noise. The full channel matrix \( \mathbf{H} \in \mathbb{C}^{N_S \times N_D} \) over subcarriers and \ac{OFDM} symbols must be estimated from a sparse set of pilot observations $\mathcal{P}$.
Classical methods such as \ac{LS} and \ac{MMSE} \cite{singh2014performance} rely on pilot-based interpolation \cite{Zhang2010} to reconstruct the full channel, but each has drawbacks. While \ac{LS} is computationally efficient, it is sensitive to noise. \ac{MMSE} provides higher accuracy by using prior channel statistics, but is computationally intensive and difficult to implement in dynamic environments. To address this, recent approaches reformulate \ac{CE} as a \ac{SR} problem, aiming to recover the full-resolution channel $\hat{\mathbf{H}}$ from sparse pilot observations $\mathbf{H}_p$, modeled as: 
\begin{equation}\label{dl_function}
    \hat{\mathbf{H}} = f_{\Theta}(\mathbf{H}_{LR}),
\end{equation}  
where $f_{\Theta}$ denotes a \ac{DL} model parameterized by $\Theta$ that learns a mapping from a low-resolution input $\mathbf{H}_{LR}$  to the full channel estimate $\hat{\mathbf{H}}$. The input $\mathbf{H}_{LR}$ can be obtained either directly from raw pilot observations (($\mathbf{H}_{LR} = \mathbf{H}_p$)) or from \ac{LS}-based estimates combined with \ac{LI}, providing a partial yet denoised representation of the true channel $\mathbf{H}$. This formulation is analogous to \ac{SR} in image processing, where high-resolution details are reconstructed from degraded or sparse inputs. The objective is to design $f{\Theta}$ such that it accurately reconstructs $\hat{\mathbf{H}}$ while meeting the low-latency requirements of real-time 5G systems.

The problem can be formalized as the following constrained optimization:
\begin{equation}\label{opt_form}
\min_\Theta \; \mathbb{E} \left[
\| f_{\Theta}(\mathbf{H}_{LR}) - \mathbf{H} \|_F^2
\right]
\quad \text{s.t.} \quad T_\mathrm{inf}(f_\Theta) \leq T_{\max},
\end{equation}
where \( \| \cdot \|_F \) is the Frobenius norm, and the objective corresponds to the \ac{MSE} between the predicted and true channel matrices. \( T_\mathrm{inf}(f_\Theta) \) denotes the model's inference time, and \( T_{\max} \) is the allowable latency budget. This formulation captures the need to jointly optimize estimation accuracy and computational efficiency for real-time 5G systems. While \ac{MSE} serves as the training objective, we report performance using the \ac{NMSE}, defined as
\begin{equation}\label{eq:nmse}
\mathrm{NMSE} = \frac{\mathbb{E}\left[\| \hat{\mathbf{H}} - \mathbf{H} \|_F^2\right]}{\mathbb{E}\left[\| \mathbf{H} \|_F^2\right]},
\end{equation}
which offers a scale-invariant measure of estimation quality across varying signal and channel conditions. All results are reported in \ac{dB}.

Prior work such as ChannelNet~\cite{soltani2019deep}, EDSR~\cite{maruyama2021}, and more recently \ac{ProEsNet}~\cite{Zhang2024}, apply deep \acp{CNN} inspired by image \ac{SR} to enhance \ac{CE} accuracy, albeit with increased latency. \ac{SRCNN}, uses as first stage of ChannelNet, reduces complexity but suffers in performance. \ac{AttRNet}~\cite{Gao2025} improves efficiency using SE blocks, and \ac{LSiDNN}~\cite{Sharma2024} explicitly targets low complexity with dense layers. \ac{EPformer}~\cite{Guo2024} and \ac{CEViT}~\cite{Liu2024} leverages Transformer-based architecture for top accuracy, but incurs high computational cost. These limitations motivate \ac{HELENA}, which combines shallow convolution, dual attention, and compact design to jointly optimize accuracy, complexity, and latency.

%% file: sections/section3.tex
\section{HELENA Architecture}
\ac{HELENA} is a lightweight \ac{DL} model for pilot-based \ac{CE} in \ac{5G-NR} \ac{OFDM} systems, designed to balance accuracy and inference time. It combines convolutional feature extraction with dual attention: patch-wise \ac{MHSA} for long-range context and a low-cost \ac{SE} block for channel-wise recalibration. The input is a sparse \ac{LS}-based estimate \(\mathbf{H}_{\text{LR}} \in \mathbb{R}^{N_S \times N_D \times 2}\), where values are present only at pilot positions and set to zero elsewhere, while the output is a full-resolution estimate \(\hat{\mathbf{H}} \in \mathbb{R}^{N_S \times N_D \times 2}\) covering the entire time-frequency grid. The following subsection outlines the main architectural components and their design rationale.
\begin{figure}[t!]
    \centering    
    \includegraphics[width=\columnwidth]{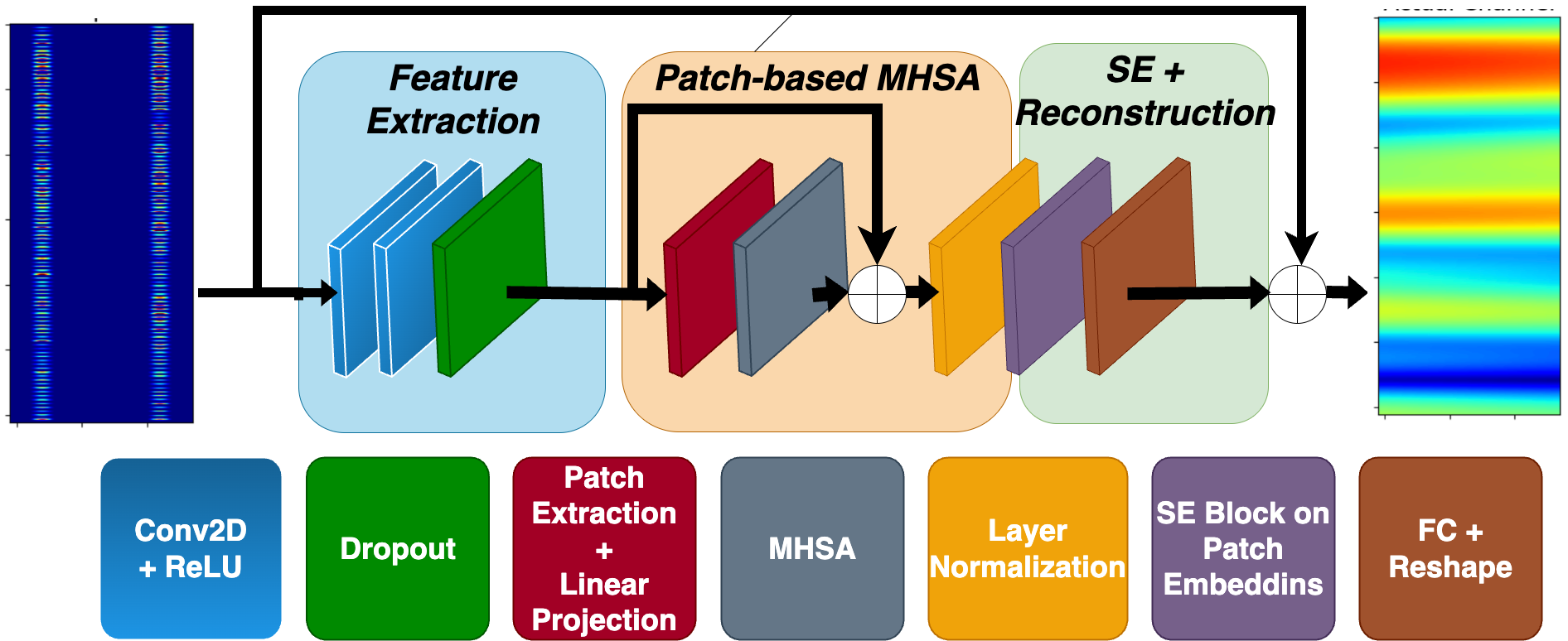}
    \caption{The proposed HELENA architecture. }
    \label{fig:helena}
\end{figure}
\subsection{Feature Extraction via Shallow CNN}
In order to extract local spatial features from \(\mathbf{H}_{\text{LR}}\), the model uses two 2D convolutional layers with ReLU activations. The first convolutional layer applies a kernel of size \(f_1 \times t_1\) and uses \(C_1\) filters, producing an intermediate feature map \(\mathbf{F}^{(1)}\). The second layer uses a kernel of size \(f_2 \times t_2\) with \(C\) filters, resulting in \(\mathbf{F}^{(2)}\). These layers have associated learnable weights \(\mathbf{W}_1, \mathbf{W}_2\) and biases \(\mathbf{b}_1, \mathbf{b}_2\), respectively. A dropout layer is applied to improve generalization:
\begin{gather}
\mathbf{F}^{(1)} = \text{ReLU}\left(\text{Conv2D}_1(\mathbf{H}_{\text{LR}}; \mathbf{W}_1, \mathbf{b}_1)\right) \\
\mathbf{F}^{(2)} = \text{ReLU}\left(\text{Conv2D}_2(\mathbf{F}^{(1)}; \mathbf{W}_2, \mathbf{b}_2)\right) \\
\mathbf{F}_{\text{drop}} = \text{Dropout}(\mathbf{F}^{(2)})
\end{gather}
The output tensor \(\mathbf{F}_{\text{drop}} \in \mathbb{R}^{N_S \times N_D \times C}\) is then partitioned into non-overlapping patches of height \(p\), giving \(N = N_S / p\) patches. This patching strategy enables localized spatial aggregation while reducing the sequence length, making the subsequent attention computation more efficient. Each patch is then flattened into a vector:
\begin{equation}
\mathbf{F}_{\text{patch}}^{(i)} \in \mathbb{R}^{p \times N_D \times C}, \quad \mathbf{P}_i = \text{Flatten}(\mathbf{F}_{\text{patch}}^{(i)})
\end{equation}
\subsection{Patch Embedding and Multi-Head Self-Attention}
Each patch vector \(\mathbf{P}_i\) is projected into a shared \(d\)-dimensional embedding space using a learnable weight matrix \(\mathbf{W}_e \in \mathbb{R}^{(p N_D C) \times d}\) and bias \(\mathbf{b}_e \in \mathbb{R}^d\):
\begin{equation}
\mathbf{Z}_i = \mathbf{P}_i \mathbf{W}_e + \mathbf{b}_e, \quad \mathbf{Z} \in \mathbb{R}^{N \times d}
\end{equation}
In \ac{HELENA}, each flattened and embedded patch vector \(\mathbf{Z}_i\), also referred to as a \emph{token}, following transformer terminology, represents a localized region of the time-frequency channel grid, corresponding to a group of neighboring subcarriers and \ac{OFDM} symbols. These patches are designed to capture local variations in the wireless channel, such as those introduced by multipath delay and Doppler shift. The resulting \(N\) tokens are stacked into a matrix \(\mathbf{Z} \in \mathbb{R}^{N \times d}\), where each row is a token embedding of dimension \(d\). This sequence is fed into the \ac{MHSA} block, which models global context and long-range dependencies in the time--frequency domain. Each patch embedding \( Z \in \mathbb{R}^{N \times d} \) is linearly projected into query, key, and value vectors, enabling the computation of attention weights that capture contextual relevance between patches. These weights guide how information from other patches is aggregated to refine each token representation. For each head \( j \), \( Z \) is projected to queries, keys, and values as:
\begin{gather}
Q_j = Z W^Q_j, \quad K_j = Z W^K_j, \quad V_j = Z W^V_j \\
\operatorname{head}_j = \operatorname{Softmax}\left( \frac{Q_j K_j^\top}{\sqrt{d_k}} \right) V_j
\end{gather}
Outputs from all heads are concatenated and projected using \( W^O \in \mathbb{R}^{h d_k \times d} \), followed by a residual connection and layer normalization to produce the final MHSA output:
\begin{gather}
\operatorname{MHSA}(Z) = \operatorname{Concat}(\operatorname{head}_1, \dots, \operatorname{head}_h) W^O \\
Z_{\text{att}} = \operatorname{LayerNorm}(Z + \operatorname{MHSA}(Z))
\end{gather}
\subsection{Channel-wise Attention and Reconstruction}
While \ac{MHSA} captures long-range dependencies and aggregates contextual information across the full time-frequency grid, it treats all embedding channels equally during its output projection. However, not all channels (i.e., feature dimensions) contribute equally to the task of \ac{CE}, e.g., some may carry more informative patterns depending on the propagation environment and pilot configuration. To recalibrate channel importance, \ac{HELENA} applies a lightweight \ac{SE} block \cite{Hu2018} across the token embeddings. First, a global descriptor \(\mathbf{s} \in \mathbb{R}^d\) is computed via average pooling:
\begin{equation}
\mathbf{s} = \frac{1}{N} \sum_{i=1}^N \mathbf{Z}_{\text{att}, i}
\end{equation}
Then, two fully connected layers with weights \(\mathbf{W}_{se1} \in \mathbb{R}^{d \times d/r}, \mathbf{W}_{se2} \in \mathbb{R}^{d/r \times d}\) and nonlinearities (ReLU and sigmoid) produce a scaled excitation vector \(\mathbf{e} \in \mathbb{R}^d\):
\begin{gather}
\mathbf{e} = \sigma\left( \mathbf{W}_{\text{se2}} \cdot \text{ReLU}(\mathbf{W}_{\text{se1}} \cdot \mathbf{s} + \mathbf{b}_{\text{se1}}) + \mathbf{b}_{\text{se2}} \right) \\
\mathbf{Z}_{\text{scaled}, i} = \mathbf{Z}_{\text{att}, i} \odot \mathbf{e}
\end{gather}
This post-attention recalibration step enables the network to focus on semantically meaningful representations by learning a global importance weighting over the embedding dimensions and selectively amplifying the most relevant features, all with minimal computational overhead. Then, each scaled token is projected back to the original patch space using reconstruction weights \(\mathbf{W}_r \in \mathbb{R}^{d \times (p N_D \cdot 2)}\) and bias \(\mathbf{b}_r \in \mathbb{R}^{p N_D \cdot 2}\), and the resulting set of patch vectors \(\{\mathbf{P}'_i\}_{i=1}^N\) is reshaped to form the final output estimate \(\hat{\mathbf{H}} \in \mathbb{R}^{N_S \times N_D \times 2}\):
\begin{gather}
\mathbf{P}'_i = \mathbf{Z}_{\text{scaled}, i} \mathbf{W}_r + \mathbf{b}_r \\
\hat{\mathbf{H}} = \text{Reshape}(\{\mathbf{P}'_i\}_{i=1}^N)
\end{gather}
Finally, a global residual connection \cite{he2016deep} is included to improve convergence and ensure preservation of coarse structural information from the initial \ac{LS} estimates \cite{Gao2025}:
\begin{equation}\label{rescon}
\hat{\mathbf{H}} = \mathbf{H}_{\text{LR}} + \text{HELENA}(\mathbf{H}_{\text{LR}})
\end{equation}
The resulting \ac{DL} architecture, \ac{HELENA} (equivalent to $f_{\Theta}$ in Eq.~\ref{dl_function}), combines local and global feature learning with low complexity. Its dual attention mechanism, \ac{MHSA} and \ac{SE}, enables accurate reconstruction without interpolation or side information, achieving high accuracy at low computational and memory cost as shown in Section~\ref{results}.

%% file: sections/section4.tex
\section{Evaluation Results}\label{results}

\begin{table}[tb]
\centering
\caption{Channel and 5G-NR Parameters for Dataset Generation}
\resizebox{\columnwidth}{!}{%
\begin{tabular}{|l|l|}
\hline
\textbf{Parameter} & \textbf{Value/Description} \\ \hline
Channel Profiles & TDL-A to TDL-E \\ \hline
Fading Distribution & Rayleigh \\ \hline
Antennas (Tx, Rx) & 1, 1 \\ \hline
Carrier Configuration & 51 \acs{RB}, 30 kHz \acs{SCS}, Normal \acs{CP} \\ \hline
Sub-carriers per \ac{RB} & 12 \\ \hline
Symbols per Slot & 14 \\ \hline
Slots per Subframe & 2 \\ \hline
Slots per Frame & 20 \\ \hline
Frame Duration & 10 ms \\ \hline
NFFT & 1024 \\ \hline
Transmission Direction & Downlink \\ \hline
\acs{PDSCH} Configuration & PRB: 0–50, All symbols, Type A, 1 layer \\ \hline
Modulation & 16QAM \\ \hline
\multirow{2}{*}{DM-RS Configuration} & Ports: 0, Type A Position: 2 \\
 & Length: 1, Config: 2 \\ \hline
Delay Spread & 1–300 ns \\ \hline
Doppler Shift & 5–400 Hz \\ \hline
\ac{SNR} & 0–20 dB (2 dB steps) \\ \hline
Sample Rate & 30.72 MHz \\ \hline
Noise & AWGN \\ \hline
Interpolation & Linear \\ \hline
\multirow{2}{*}{Dataset Shape} & [11264, 612, 14, 2] = [samples, \acs{SCS}$\times$\acs{RB}, \\
 & symbols, real \& imaginary components] \\ \hline
\end{tabular}
}
\label{table:dataset}
\end{table}

\subsection{Dataset Description}
The dataset was synthetically generated using MATLAB’s 5G Deep Learning Data Synthesis framework\footnote{https://nl.mathworks.com/help/5g/ug/deep-learning-data-synthesis-for-5g-channel-estimation.html}, with minor modifications, and emulates pilot-based \ac{CE} in realistic \ac{OFDM} systems. Key parameters, including \ac{PDSCH} configuration, carrier frequency, \ac{SCS}, \ac{CP} type, number of \acp{RB}, code rate, and modulation, are summarized in Table~\ref{table:dataset}. The \ac{SNR} range was extended to 0–20,dB (11 values, 1024 samples each), yielding 11{,}264 samples split into 70\% training, 15\% validation, and 15\% testing. Ground truth labels are derived from full \ac{CSI}, with models' input varying by method: \ac{HELENA}, \ac{LSiDNN}, and \ac{ProEsNet} use \ac{LS} estimates at pilot positions, while ChannelNet, \ac{EDSR}, \ac{AttRNet}, and \ac{CEViT} use \ac{LI}-interpolated LS. The dataset includes 3GPP TDL-A–E profiles and a range of Doppler shifts and SNR levels, enabling robust generalization across diverse propagation conditions.



\subsection{Baseline Methods, Models and Experimental Setup}

We compare \ac{HELENA} against seven \ac{DL}-based baselines, \ac{SRCNN}, ChannelNet~\cite{soltani2019deep}, \ac{EDSR}~\cite{maruyama2021}, \ac{ProEsNet}\cite{Zhang2024}, \ac{AttRNet}~\cite{Gao2025}, \ac{LSiDNN}~\cite{Sharma2024}, and \ac{CEViT}~\cite{Liu2024}, as well as two classical estimators: the \ac{LS} method with \ac{LI}, and a practical \ac{5G-NR} variant with denoising and averaging\footnote{\url{https://www.mathworks.com/help/5g/ref/nrchannelestimate.html}}\footnote{\url{https://github.com/srsran/srsRAN_Project/blob/main/lib/phy/upper/signal_processors/port_channel_estimator_average_impl.cpp}}.

All models were reimplemented with adaptations for deployment and fair comparison. ChannelNet uses 32 \ac{DnCNN} filters (vs. 64). \ac{EDSR} includes 32 filters and 16 residual blocks, restructured for joint I/Q input using 2D-\ac{CNN}s. \ac{ProEsNet} omits upsampling blocks from the original design, as it operates directly on the full resolution resource grid without requiring interpolation from sparse pilot positions. AttRNet uses the AttResNet-Conv variant (32 filters). \ac{LSiDNN} is evaluated with 48 and 1024 neurons to reflect input size. \ac{CEViT} uses \texttt{Conv2DTranspose} in place of inverse patch embedding for TensorRT compatibility. All models use \texttt{padding='same'} and omit upsampling. We also include HELENA-MHSA (without \ac{SE}) to isolate the contribution of dual attention.

Experiments were run on the GPULab testbed\footnote{\url{https://doc.ilabt.imec.be/ilabt/gpulab/}} using 4 vCPUs, 64~GB RAM, and an NVIDIA Tesla V100-SXM3 (32~GB). The software stack included CUDA~12.2, TensorFlow~2.15, and TensorRT~8.6.1. Training used Adam optimizer and \ac{MSE} loss (batch size 64), with early stopping (patience 50) and learning rate 0.01, reduced by 0.8 every 40 epochs without improvement (min $1\times10^{-5}$). Checkpoints were selected based on validation loss. Inference time was averaged over 100 single-sample runs. Results refer to TensorRT-optimized models, as non-optimized versions showed similar accuracy but are unsuitable for deployment~\cite{goez2024latincom}.
\subsection{Design Decisions and Parameter Selection in HELENA}
\begin{figure}[tb]
    \centering
    \includegraphics[width=\columnwidth]{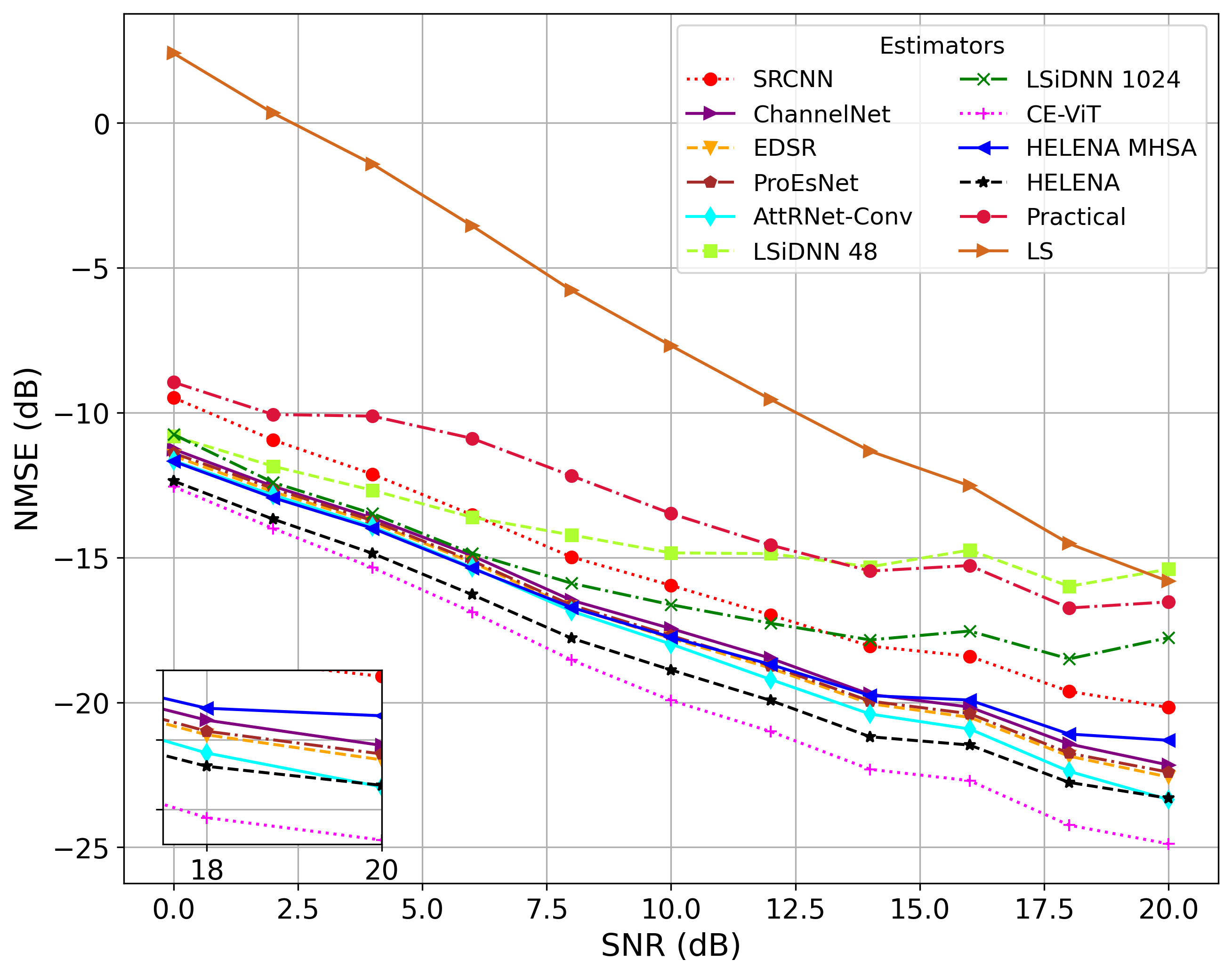}
    \caption{NMSE vs. SNR for various CE methods.}
    \label{fig:mse-vs-snr}
\end{figure}
\begin{table}[tb]
\centering
\caption{Comparison of NMSE (dB), FLOPS, and inference time using HELENA as baseline. Lower values are better.}
\renewcommand{\arraystretch}{1.25}
\setlength{\tabcolsep}{4pt}
\resizebox{\columnwidth}{!}{%
\begin{tabular}{|l|c|c|cc|cc|}
\hline
\textbf{Model} & 
\begin{tabular}[c]{@{}c@{}}\textbf{Params}\\ ($\times 10^6$)\end{tabular} & 
\begin{tabular}[c]{@{}c@{}}\textbf{FLOPS}\\ ($\times 10^9$)\end{tabular} &
\multicolumn{2}{c|}{\textbf{NMSE (dB)}} &
\multicolumn{2}{c|}{\textbf{$T_\mathrm{inf}$ (ms)}} \\
\cline{4-7}
 & & & Value & $\Delta$ & Value & $\Delta$ \\
\hline
SRCNN         & 0.014 & 0.241  & -13.828 & \cellcolor{red!20}17.60\% & 0.120 & \cellcolor{green!20}-31.43\% \\
ChannelNet    & 0.184 & 3.108  & -15.507 & \cellcolor{red!20}7.59\%  & 0.293 & \cellcolor{red!20}67.43\% \\
EDSR          & 0.306 & 5.245  & -15.773 & \cellcolor{red!20}6.03\%  & 0.388 & \cellcolor{red!20}121.71\% \\
ProEsNet      & 0.170 & 2.915  & -15.684 & \cellcolor{red!20}6.54\%  & 0.466 & \cellcolor{red!20}166.29\% \\
AttRNet-Conv  & 0.075 & 1.288  & -15.993 & \cellcolor{red!20}4.70\%  & 0.293 & \cellcolor{red!20}67.43\% \\
LSiDNN 48     & 1.662 & 0.003  & -13.546 & \cellcolor{red!20}19.28\% & 0.0738 & \cellcolor{green!20}-57.83\% \\
LSiDNN 1024   & 35.11 & 0.070  & -14.834 & \cellcolor{red!20}11.61\% & 0.158 & \cellcolor{green!20}-9.71\% \\
CE-ViT        & 0.880 & 0.053  & -17.303 & \cellcolor{green!20}-3.11\% & 0.318 & \cellcolor{red!20}81.71\% \\
HELENA MHSA   & 0.114 & 0.072  & -15.839 & \cellcolor{red!20}5.62\%  & 0.172 & \cellcolor{green!20}-1.71\% \\
\rowcolor{blue!10}
HELENA        & 0.116 & 0.077  & -16.782 & ---      & 0.175 & --- \\
\hline
\end{tabular}
}
\label{tab:nmse_flops_time_compact}
\end{table}
The parameters of \ac{HELENA} were chosen based on the structural characteristics of the input data and validated empirically to optimize the trade-offs between estimation accuracy, inference time, and model complexity. The convolutional layers use \(12 \times 2\) and \(6 \times 7\) kernels to exploit the time-frequency layout of the \ac{OFDM} grid while remaining lightweight. A patch size of 12 along the frequency axis yields 51 tokens, matching the 51 \acp{RB} in the dataset. Each token is embedded into a 64-dimensional space, which offers sufficient capacity for expressive representation while avoiding overfitting or excessive compute. The \ac{MHSA} block uses 4 heads, allowing parallel modeling of diverse attention patterns without excessive overhead while maintaining high accuracy. A reduction ratio of 4 in the \ac{SE} block avoids aggressive bottlenecking, enabling effective channel-wise recalibration at low cost. Dropout (0.1) and residual connections further enhance generalization and convergence.
\subsection{Model Accuracy}
Fig.~\ref{fig:mse-vs-snr} shows the variation of \ac{NMSE}, as defined in Eq.~\ref{eq:nmse}, across different \ac{SNR} levels. \ac{CEViT} achieves the best NMSE ($-17.303$~dB), followed closely by \ac{HELENA} ($-16.782$~dB), which requires no explicit \ac{SNR}, Doppler, or latency inputs and avoids interpolation. Compared to \ac{LS} ($-3.56$~dB) and the practical estimator ($-12.25$~dB), \ac{HELENA} improves \ac{NMSE} by 95.3\% and 65.5\%, respectively. Relative to other \ac{DL} baselines, \ac{HELENA} improves over ProEsNet by $6.54\%$, EDSR by 6.03\%, AttRNet by 4.70\%, and ChannelNet by 7.59\%. Compared to the lighter models, the gain is even higher: 17.60\% over SRCNN, 19.28\% over LSiDNN-48, and 11.61\% over LSiDNN-1024, highlighting the trade-off between model complexity and diminishing performance returns. Finally, removing the \ac{SE} block degrades accuracy by 5.62\%, confirming the benefit of dual attention, yet the model still outperforms all lighter baselines and the most accurate \ac{SR}-based models (ProEsNet, EDSR, ChannelNet), in average across all \acp{SNR}.
\subsection{Computational and Memory Cost vs. Inference Time}\label{inference_analysis}
To meet the stringent latency requirements of \ac{5G-NR}, \ac{CE}, equalization, and decoding must complete within the \ac{HARQ} deadline, which allows up to three \acp{TTI} (0.5~ms per \ac{TTI} at 30~kHz SCS)~\cite{Damjancevic2021}. This implies a tight budget $T_{\max}$ of roughly 0.5~ms for \ac{CE}. As shown in Table~\ref{tab:nmse_flops_time_compact}, all models satisfy the latency constraint \( T_\mathrm{inf}(f_\Theta) \leq T_{\max} \) from Eq.~\ref{opt_form}, albeit with varying margins.

\ac{HELENA} achieves a strong trade-off: it delivers the second-best accuracy, just 3.1\% lower than \ac{CEViT}, and has the lowest inference time among the top-performing models (excluding HELENA MHSA). Specifically, \ac{CEViT} and \ac{AttRNet} are 81.71\% and 67.43\% slower, respectively (0.318 ms and 0.293 ms vs. 0.175 ms). Compared to the lighter models, \ac{HELENA} is 31.43\% slower than SRCNN (0.120ms) and LSiDNN-48 (0.0738ms, 57.83\% fastest) but they achieve it at the cost of higher prediction error. HELENA MHSA, without the \ac{SE} block, performs similarly at 0.172ms. ProEsNet, while achieving comparable accuracy to ChannelNet and EDSR, exhibits the highest latency among top models (0.466ms), consuming over 90\% of the latency budget. In general, while all models satisfy Eq.~\ref{opt_form}, \ac{HELENA} uses only 35\%, highlighting its suitability for real-time deployment.

We can see that runtime does not scale linearly with computational complexity (in \ac{FLOPS}\footnote{Measured using TensorFlow’s profiling tool: \url{https://www.tensorflow.org/api_docs/python/tf/compat/v1/profiler/ProfileOptionBuilder}}). For example, \ac{HELENA} uses 66.7\% fewer \ac{FLOPS} than \ac{SRCNN} (0.08 vs. 0.24 G), yet runs 38.5\% slower. AttRNet and EDSR require 16× and 65× more operations than \ac{HELENA}, but inference is only 65.6\% and 96.5\% slower, respectively. ChannelNet also runs 37.9\% slower, despite requiring 39× more \ac{FLOPS}. In contrast, ProEsNet performs 2.9 GFLOPS per inference, over 36× more than HELENA, yet is only 2.7× slower, reinforcing the decoupling between FLOPS and latency due to implementation and hardware bottlenecks.

The number of parameters, which relates to the memory required at inference time, shows limited correlation with runtime or estimation accuracy. \ac{HELENA} reaches $-16.87$~dB NMSE with only 0.11M parameters, outperforming ChannelNet (0.18M) and LSiDNN-1024 (35.11M) while being faster. \ac{CEViT} achieves the best NMSE with 0.88M parameters, 8× more than \ac{HELENA}, but is slower. ProEsNet has a moderate footprint of 0.17M parameters. These results emphasize the need to jointly consider \ac{FLOPS}, parameter count, and inference time when optimizing models for real-time \ac{CE}, as actual latency is shaped by hardware–software co-design aspects such as memory access, operator fusion, and backend scheduling, which extend beyond theoretical complexity.

%% file: sections/section5.tex
\section{Conclusion and Future Works}

This paper presented \ac{HELENA}, a \ac{DL}-based model for efficient and accurate pilot-based \ac{CE} in \ac{OFDM} systems. HELENA integrates shallow convolutional layers with dual attention—patch-wise \ac{MHSA} and channel-wise \ac{SE}—to balance estimation accuracy, computational complexity, and inference latency. Experimental results demonstrate that HELENA achieves near state-of-the-art performance with low computational and memory requirements while maintaining fast inference, making it well suited for real-time, latency-sensitive wireless systems. The observed weak correlation between \ac{FLOPS}, parameter count, and actual inference time further highlights the importance of holistic evaluation criteria when designing practical \ac{DL}-based \ac{CE} models. Although HELENA builds upon established architectural components, this work shows that a careful and lightweight combination of attention mechanisms can yield deployment-ready performance under strict latency constraints.

Future research will explore a broader architectural design space through systematic ablation studies, including the impact of patch size, embedding dimension, and the number of attention heads, in order to better characterize accuracy-complexity trade-offs. In addition, although the current evaluation relies on synthetically generated datasets, validating HELENA using real wireless measurements and hardware testbeds remains an important next step to assess robustness under practical RF impairments. The framework is also expected to be extended to \ac{MIMO}, massive \ac{MIMO}, and beamformed systems, which will require adaptations to efficiently exploit spatial correlations across antennas. Further directions include the visualization and analysis of attention maps to improve model interpretability, as well as the incorporation of statistical performance indicators such as variance and confidence intervals alongside average \ac{NMSE}. Finally, deployment on hardware-constrained platforms, including \acs{FPGA} and edge \ac{AI} accelerators, will be investigated to further evaluate real-time feasibility.